\documentclass[twocolumn,superscriptaddress,amsmath,amssymb,showpacs,prl]{revtex4-1}

\usepackage{graphicx}
\usepackage{color}
\renewcommand{\phi}{\varphi}

\begin{document}

\title{Predicting Complex Relaxation Processes in Metallic Glass}

\author{Yang Sun}
	\thanks{These authors contributed equally.}
	\affiliation{Ames Laboratory, US Department of Energy and Department of Physics, Iowa State University, Ames, Iowa 50011, USA}
\author{Si-Xu Peng}
	\thanks{These authors contributed equally.}
	\affiliation{Wuhan National High Magnetic Field Center and School of Physics, Huazhong University of Science and Technology, Wuhan, Hubei 430074, China}
\author{Qun Yang}
		\affiliation{Wuhan National High Magnetic Field Center and School of Physics, Huazhong University of Science and Technology, Wuhan, Hubei 430074, China}
\author{Feng Zhang}
	\affiliation{Ames Laboratory, US Department of Energy and Department of Physics, Iowa State University, Ames, Iowa 50011, USA}
\author{Meng-Hao Yang}
	\affiliation{Ames Laboratory, US Department of Energy and Department of Physics, Iowa State University, Ames, Iowa 50011, USA}
\author{Cai-Zhuang Wang}
	\affiliation{Ames Laboratory, US Department of Energy and Department of Physics, Iowa State University, Ames, Iowa 50011, USA}
\author{Kai-Ming Ho}
	\affiliation{Ames Laboratory, US Department of Energy and Department of Physics, Iowa State University, Ames, Iowa 50011, USA}
\author{Hai-Bin Yu}
	\email[Email: ]{haibinyu@hust.edu.cn}
	\affiliation{Wuhan National High Magnetic Field Center and School of Physics, Huazhong University of Science and Technology, Wuhan, Hubei 430074, China}

\begin{abstract}

Relaxation processes significantly influence the properties of glass materials. However, understanding their specific origins is difficult, even more challenging is to forecast them theoretically. In this study, using microseconds molecular dynamics simulations together with an accurate many-body interaction potential, we predict that an Al$_{\text{90}}$Sm$_{\text{10}}$ metallic glass would have complex relaxation behaviors: In addition to the main ($\alpha$) relaxation, the glass (i) shows a pronounced secondary ($\beta$) relaxation at cryogenic temperatures and (ii) exhibits an anomalous relaxation process ($\alpha_2$) accompanying $\alpha$ relaxation. Both of the predictions are verified by experiments. Computational simulations reveal the microscopic origins of relaxation processes: while the pronounced $\beta$ relaxation is attributed to the abundance of string-like cooperative atomic rearrangements, the anomalous $\alpha_2$ process is found to correlate with the decoupling of the faster motions of Al with slower Sm atoms. The combination of simulations and experiments represents a first glimpse of what may become a predictive routine and integral step for glass physics.

\end{abstract}

\maketitle

Compared with crystals, glasses inherently feature diverse relaxation dynamics over a wide range of temperature and timescales. These relaxation processes significantly influence properties of glass materials and are related to a number of crucial unresolved issues in glassy physics ~\cite{Berthier2011, Angell2000, Micoulaut2016, Lunkenheimer2000,Sokolov2015}. Understanding how the atomic rearrangements govern these processes represents an outstanding issue in glass physics \cite{Berthier2011, Micoulaut2016, Pazmi2015, Lunkenheimer2000, Lahini2017}.

Usually, the most prominent relaxation process is the primary ($\alpha$) relaxation which is responsible for the vitrification of glass-forming liquid. Its falling out of equilibrium  indicates the glass transition phenomenon. Processes occurring in addition to the $\alpha$ relaxation at shorter timescales or lower temperature are referred as secondary ($\beta$) relaxations. The studies over the last a few decades have established that the $\beta$ relaxation could have important consequences on the mechanical properties for metallic and polymeric glasses, as well as thermal stability of glassy pharmaceuticals and biomaterials, and thus attract considerable attentions \cite{Yu2014, Yu2013, Ngai2004, Capaccioli2012, Geirhos2018, Sondhaub2015}. Moreover, recent studies have discovered there might be more relaxation processes in addition to the $\alpha$ and $\beta$ relaxations in glasses \cite{Lahini2017, Welch2013,  Kuchemann2017, Luo2017, Wang2015, Yu2017, Cangialosi2013, Ruta2012,Bi2018}. Even the structurally simplest metallic glasses (MGs, compared to molecular and polymeric glasses) could exhibit multiple relaxations, indicating that a far richer-than-expected scenario for glass relaxation. For example, Wang \textit{et al.} \cite{Wang2015} and Küchemann \textit{et al.} \cite{Kuchemann2017}  identified a new relaxation process that is faster than the $\beta$ relaxation in MGs which was named  $\beta'$ and $\gamma$ relaxations. Wang \textit{et al.}  \cite{Wang2017} also illustrate the $\beta'$ relaxation might be correlated with the initiation of plastic deformation in MGs. Luo and coworkers \cite{Luo2017} reported the non-equilibrium $\alpha$ relaxation would split into two processes in deep glassy state which causes early decay in the stress-relaxation experiments \cite{Ruta2017}. While these results illustrate that MGs possess complex relaxation phenomena and have consequences on properties, it is difficult to understand their specific origins because of the lack of microscopic data of these processes. 

In principle, molecular dynamics (MD) simulation is a powerful tool to investigate detailed atomic rearrangements in the glass relaxations at the microscopic level. However, the relaxations in the glassy states are extremely complicated, as they are sensitive to chemical compositions \cite{Yu2013a} and thermal histories \cite{Giordano2016, Wang2016}. The related relaxation timescales (e.g., milliseconds to seconds) are usually several decades longer than the current available computational timescales (picoseconds to nanoseconds). It is therefore challenging to model the relaxation dynamics of realistic glass materials under experimental conditions. Moreover, for large-scale MD simulations of MGs, the force field which describes the many-body interactions (i.e., empirical potentials) is of vital importance \cite{Cheng2009, Cheng2011, Falk2019}. Although there are a few potentials that can reproduce some static structural features and thermodynamics of MGs, their capability of describing dynamical processes is mostly unknown. 

Recently, a realistic interaction potential for the study of  Al$_{\text{90}}$Sm$_{\text{10}}$ MG was developed \cite{Mendelev2015}. It correctly describes the glass structure \cite{Sun2016}, complex devitrification behaviors \cite{Ye2017, Yang2018} and crystal growth \cite{Wang2017a} in the Al-Sm systems. It also brings insights to the competition of crystallization and glass formation \cite{Sun2019, Zhao2018}. Therefore, it provides a model system to investigate the relaxation mechanism in the realistic MG.

In this work, relying on this accurate interaction potential, we simulate the dynamical mechanical spectroscopy (DMS) of Al$_{\text{90}}$Sm$_{\text{10}}$ MG in the timescale up to 10 microseconds which almost reaches the limit of state-of-the-art computational power. With such a slow frequency, we find an anomalous relaxation process (noted as $\alpha_2$) decouples from $\alpha$ relaxation. The MG also exhibit a strongly pronounced $\beta$ relaxation even on the MD timescale. The behaviors predicted from the MD simulations are verified with DMS experiments at cryogenic temperatures. The detailed atomic motions that lead to the relaxation processes are revealed from the MD trajectories. The feasibility that atomic simulations could discover new relaxation processes in MGs and elucidate their underlying mechanisms would be useful for understanding the dynamics and properties as well as the design of glass materials. 

\textit{Simulation} - MD simulations were carried out based on a Finnis-Sinclair potential \cite{Mendelev2015}, using the GPU-version of LAMMPS code \cite{Brown11,Brown12,Brown13}. The Al$_{\text{90}}$Sm$_{\text{10}}$ glass model, containing 32,000 atoms, were obtained by the continuous cooling with a cooling rate of $10^{8}$K/s. The MD simulations of DMS (MD-DMS) \cite{Yu2014a} were performed during the cooling process, covering a wide temperature range from deeply undercooled liquid to low-temperature glass. In MD-DMS, a sinusoidal strain was applied with an oscillation period $t_\omega$ (related to frequency $f=1/t_\omega$) and a strain amplitude $\varepsilon_A$. The resulted stress $\sigma(t)$ and phase difference $\delta$ between stress and strain were measured and fitted by 
$ \sigma (t) = \sigma_0 + \sigma_{A} \text{sin} ( 2 \pi t / t_{\omega} + \delta) $. The storage and loss moduli were calculated by 
$E' = \sigma_A / \varepsilon_A \text{cos} (\delta)$ and 
$E'' = \sigma_A / \varepsilon_A \text{sin} (\delta)$, 
respectively. A strain amplitude $\varepsilon_A=0.6 \%$ was applied in all MD-DMS, which ensured deformations in the linear response regime.

\textit{Experiments} - The Al$_{\text{90}}$Sm$_{\text{10}}$ MG was prepared by spinning-quenching techniques (see Supplemental Material \cite{Supp} for details). The relaxation dynamics of the MG was studied by a dynamical mechanical analyzer using liquid Nitrogen for temperature control, which allows us to reach the cryogenic temperature (down to 150K). The measurements were conducted during a temperature ramping of 3 K/min together with a film tension oscillation using the discrete testing frequencies of 0.5, 2, 4, 8 and 16 Hz. The storage ($E'$) and loss ($E''$) moduli were recorded for subsequent analysis and comparing with MD-DMS results.

\begin{figure}[t]
\includegraphics[width=0.51\textwidth]{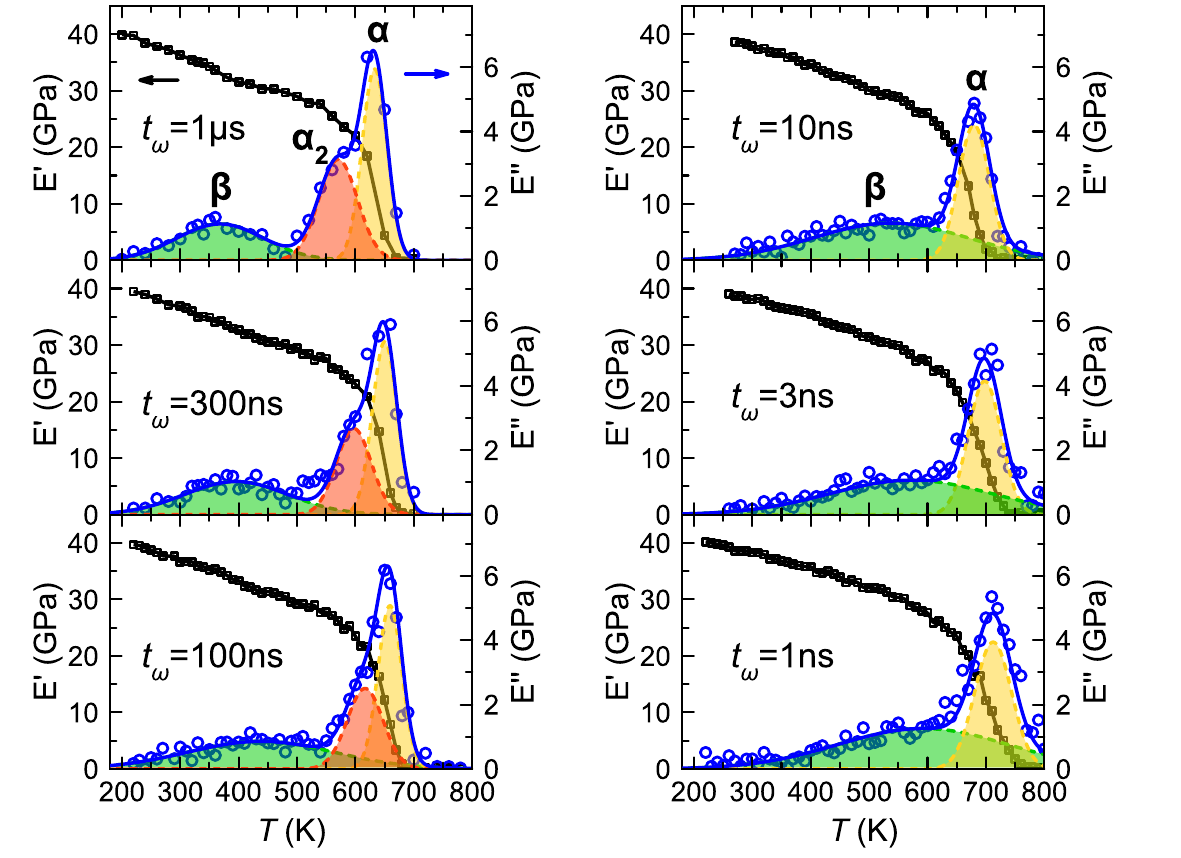}
\caption{\label{fig:fig1} Relaxation behaviors by simulations. Storage $E'$ and loss $E''$ moduli of Al$_{\text{90}}$Sm$_{\text{10}}$ MG at different oscillations periods $t_\omega$ ranging from 1 $\mu s$ to 1 $ns$. The loss moduli are fitted by multiple Gaussian functions. }
\end{figure}

\textit{Predictions by simulations} - Figure~\ref{fig:fig1} shows MD-DMS results by plotting the storage ($E'$) and loss moduli ($E''$) as a function of temperature under different oscillation periods. The temperature-dependent loss modulus curves are fitted with a serial of Gaussian peaks which correspond to different relaxation processes. With the longest oscillation period $t_\omega=1 \mu s$ in Fig.~\ref{fig:fig1}, the loss modulus exhibits a broad peak in the temperature range from 200 K to 500 K, which corresponds to the typical $\beta$ relaxation. We note that there exists no such pronounced peak of $\beta$ relaxation in any previous MD simulations where only shoulder-like or excess wings were observed. At higher temperature, the dominant primary ($\alpha$) relaxation peak shows a strong shoulder in the temperature range 500-600 K, which needs an extra peak function for the fitting. Considering this process decreases in amplitude as frequency increases (or $t_\omega$ decrease), which is consistent with behaviors of $\alpha$ relaxation in general, this new peak is named as $\alpha_2$ process.

As shown in Fig.~\ref{fig:fig1}, with the decrease of oscillation period, all three peaks, $\alpha$, $\alpha_2$ and $\beta$, shift towards higher temperature. The $\alpha_2$  peak gradually collapses in the $\alpha$  peak so that one can hardly differentiate them when $t_\omega$ is smaller than $10 ns$. With very short oscillation period (e.g. $t_\omega = 1 ns$), the $\beta$ peak also largely overlaps with the $\alpha$ peak. Thus, the simulations predict a complex relaxation scenario in the Al$_{\text{90}}$Sm$_{\text{10}}$ MG: at suitably long  time scales (e.g., $1 \mu s$) it has a pronounced $\beta$ peak and an anomalous $\alpha_2$ process in addition to the $\alpha$ relaxation.

\begin{figure}
\includegraphics[width=0.45\textwidth]{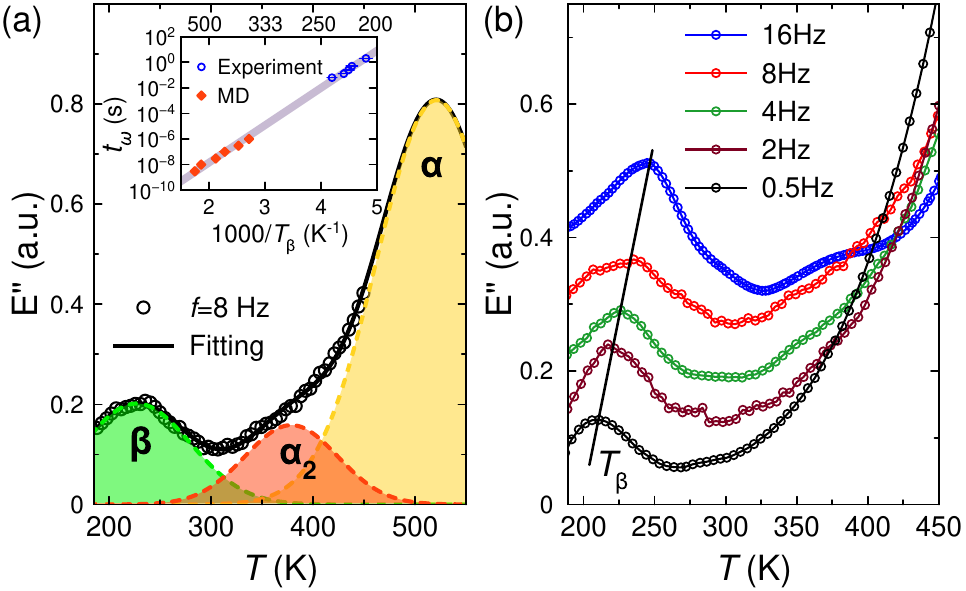}
\caption{\label{fig:fig2} Experimental DMS for Al$_{\text{90}}$Sm$_{\text{10}}$ MG. (a) DMS measured at the frequency 8 Hz and (b) at different frequencies as indicated. The inset in (a) shows the Arrhenius fitting of $\beta$ relaxation peaks and DMS periods from both simulations and experiments. The fitting of $\alpha$ relaxation peak is to guide eyes. It is based on the extrapolation of simulation data at the experimental temperature, see details in Supplemental Material \cite{Supp}.} 
\end{figure}

\textit{Experimental verification} - We next validate these computational predictions by DMS experiments for the as-quenched Al$_{\text{90}}$Sm$_{\text{10}}$ samples. Even though our longest simulation period reaches 1 $\mu s$, it is still about 5-6 orders of magnitude faster than the typical DMS experiments in which probing timescales are $0.1-10 s$. Extrapolating the temperature-time relation for the $\beta$ relaxation from simulations to the experimental timescale leads to a characteristic temperature about 200 K (inset of Fig.~\ref{fig:fig2}(a)). 

Figure~\ref{fig:fig2}(a) shows the experimental $E''(T)$ for the MG from a cryogenic DMS measurement at the testing frequency of 8 Hz. Remarkably, it clearly shows a pronounced $\beta$ relaxation peak at about 220 K, consistent with the extrapolation of MD simulations. Figure~\ref{fig:fig2}(b) shows $E''(T)$ for more different testing frequencies. The peak temperature of the $\beta$ relaxation increases with higher frequency (or shorter time period), which is also quantitatively agree with the MD simulations as shown in the inset of Fig.~\ref{fig:fig2}(a). 

Besides, a close examination of $E''(T)$ curve in Fig.~\ref{fig:fig2}(a) indicates that there is notable excess contribution to the $\alpha$ relaxation around $T = 380 K$, which is corresponding well with the $\alpha_2$ relaxation found from MD-DMS (Fig.~\ref{fig:fig1}). This $\alpha_2$ process can also be discerned from different testing frequencies in Fig.~\ref{fig:fig2}(b). For example, it is more evident on the curve with the frequency of 16 Hz. Due to close coupling with $\alpha$ relaxation, $\alpha_2$ process is more difficult to be resolved than the pronounced $\beta$ relaxation. Nevertheless, its time-temperature relation can still be determined and compared favorably with MD simulations (see Supplemental Material \cite{Supp}). Therefore, the presence of $\alpha_2$ process in the MG could be ascertained with the guidance of MD simulation. Unfortunately, because of the occurrence of strong devitrification process in the current MG system \cite{Ye2017} , one cannot fully access the $\alpha$ relaxation peak at experimental timescales, resulting in the termination of experimental data at 450K.

\textit{Mechanism for $\beta$ peak} - The above experiments validate the predictions from MD simulations. Now we are in position to investigate the mechanisms of these complex relaxation processes. Recently, the structural rearrangements governing the $\beta$ relaxations have been investigated in a model Ni$_{\text{80}}$P$_{\text{20}}$ MG \cite{Yu2017a} and a Y$_{\text{65}}$Cu$_{\text{35}}$ MG \cite{Yu2018}, which suggests string-like motions might be the origin of $\beta$ relaxation. However, these MGs do not show such clear $\beta$ relaxation peak as the Al$_{\text{90}}$Sm$_{\text{10}}$ at MD accessible timescales. One feature about string-like motions is that a particle jumps to a position that was occupied previously by another particle \cite{Yu2017a}. Structurally, this would result in a multi-peak curve for the distribution $p(u)$ of atomic displacements $u$ during the deformation period, which is clearly observed in the present MG as shown in Fig.~\ref{fig:fig3}(a) (the mathematical definitions of displacement $u$ and string-like motion are provided in Supplemental Material S3 \cite{Supp}). The fact that the second and third peaks of $p(u)$ match exactly the first and second peaks of pair distribution function $g(r)$ at various $\beta$ relaxation peaks in Fig.~\ref{fig:fig3}(a) indicates that atoms prefer to jumping to the position that is previously taken by another atom in its nearest or secondary neighbors, which further evidences string-like motions. Figure~\ref{fig:fig3}(b) shows that the strings can propagate in a rather large spatial range and form different types of geometries such as aggregations, loops and long-range chains. The string size $\xi$ is defined by the number of atoms involved in the string. Figure~\ref{fig:fig3}(c) shows that the probability of the atom forming string follows an exponential function with the string size that can span up-to 70 atoms. While similar string-like motions were also observed in Lennard-Jones liquid model \cite{Donati1998} and other systems \cite{starr2013, Zhang2015}, the string size in the current Al$_{\text{90}}$Sm$_{\text{10}}$  is much longer than other systems. For example, the longest reported string in Ni$_{\text{80}}$P$_{\text{20}}$ contains 12 atoms \cite{Yu2018}  which is smaller by a factor 6 than current MG.
Such long-range and large-scale string-like motion is a unique feature of the present MG which could be the reason for the uniquely pronounced $\beta$ peak. 

\begin{figure}
\includegraphics[width=0.48\textwidth]{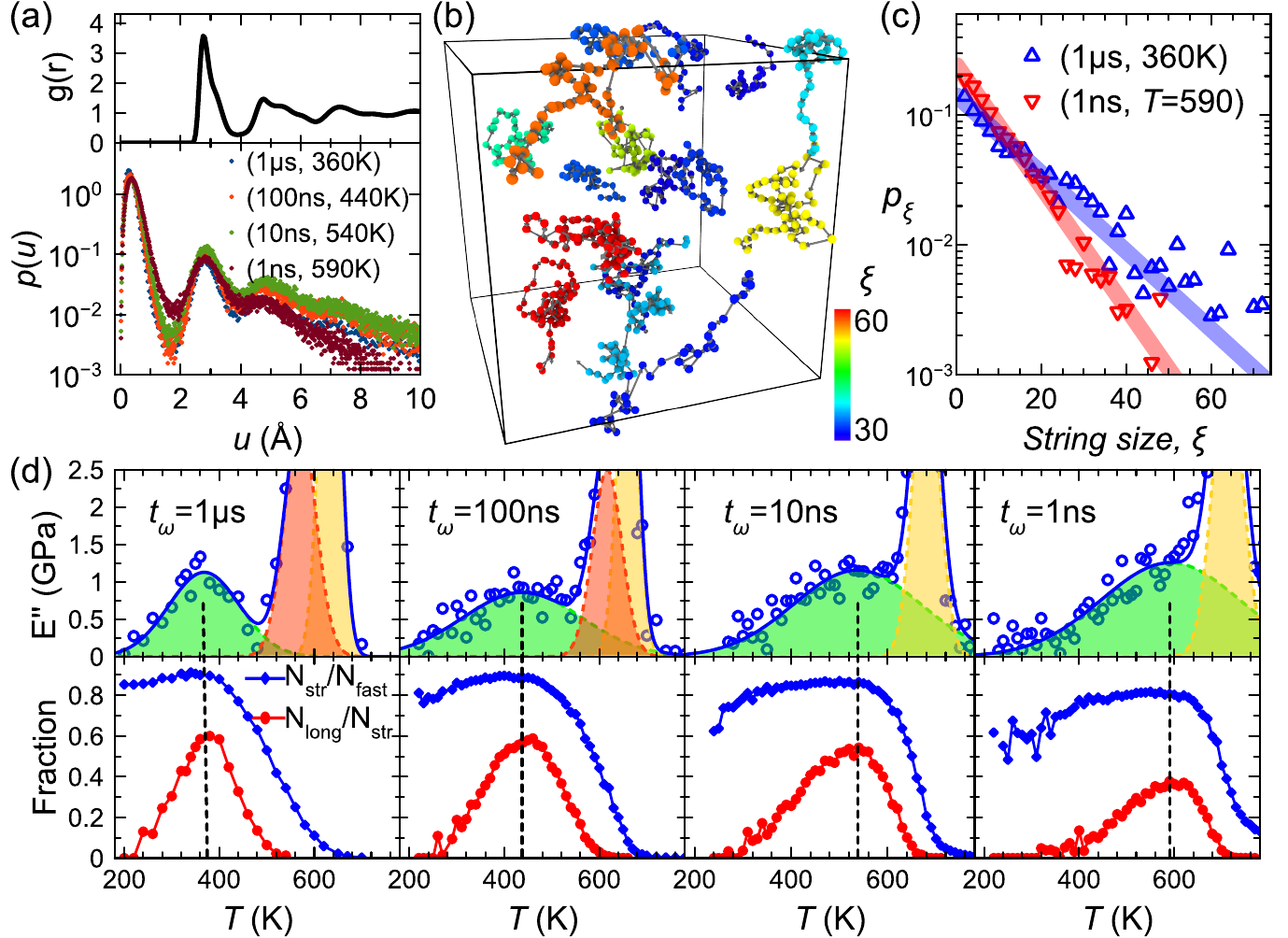}
\caption{\label{fig:fig3} String-like motion and $\beta$  relaxation. (a) Upper panel shows the pair correlation function of Al$_{\text{90}}$Sm$_{\text{10}}$ MG at 300 K from MD simulation, while lower panel shows the probability of atomic displacement $u$ at the condition of  ($t_\omega$, $T_\beta$) , where $T_\beta$ is the peak temperature of $\beta$ relaxation with the oscillation period $t_\omega$. (b) String configurations at $T_\beta  = 360 K$ with $t_\omega =  1 \mu s$. The atoms in the strings with size $\xi < 30$ are removed for clarity. (c) The probability of string-like-moving atoms involved in the different sizes of strings at $T_\beta$ with $t_\omega =  1 ns$ and $1 \mu s$, respectively. The lines indicate exponential fitting. (d) Relationship between $\beta$ relaxation and string motion under four different oscillation periods from $1 \mu s$ to $1 ns$. The lower panel shows the fraction of string-like-moving atoms in the fast-moving atoms ($\text{N}_{\text{str}} / \text{N}_{ \text{fast} } $, blue), and the fraction of atoms in the long string-like motions in the total string-like-moving atoms ($\text{N}_{\text{long}} / \text{N}_{ \text{str} } $, red). }
\end{figure}

Figure~\ref{fig:fig3}(d) quantifies the fraction of string-like-moving atoms to the total number of fast-moving atoms ($\text{N}_{\text{str}} / \text{N}_{ \text{fast} } $), as well as the fraction of atoms involved in the long-string motions to the total number of string-like-moving atom ($\text{N}_{\text{long}} / \text{N}_{ \text{str} } $) as a function of temperature. The long-string motion is defined by $\xi \geq 10$ here. The $T$-dependent loss moduli $E''$ are also plotted for comparison. One can see the peak of $\text{N}_{\text{long}} / \text{N}_{ \text{str} } $ well matches the peak of $\beta$ relaxation over all the studied oscillation periods $t_\omega$. While the curve of $\text{N}_{\text{str}} / \text{N}_{ \text{fast} } $ reaches a maximum plateau at the peak temperature of $\beta$ relaxation $T_\beta$, $\text{N}_{\text{long}} / \text{N}_{ \text{str} } $ manifests as a pronounced peak, whose position and width quantitatively agree with those of $\beta$ relaxation peaks. Note this correlation does not change with the choice of long-string motion threshold (see Supplemental Material \cite{Supp}). These results suggest long-string motions contribute more to the $\beta$ relaxation than shorter ones, which emphasis the cooperative nature of $\beta$ relaxation.

\begin{figure}
\includegraphics[width=0.49\textwidth]{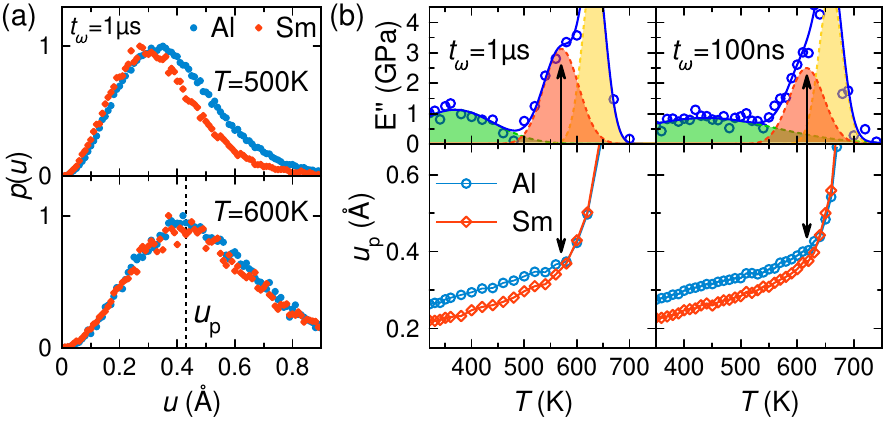}
\caption{\label{fig:fig4} Atomic motion decoupling and $\alpha_2$ relaxation. (a) The probability of Al and Sm displacement with the oscillation period $t_\omega = 1 \mu s$ at 500 K and 600 K, respectively. The dashed line indicates the peak position $u_p$. (b) $u_p$ of Al and Sm atoms and loss moduli as a function of temperature for $t_\omega = 1 \mu s$ and $100 ns$, respectively. The arrows highlight the transition points and the peak positions of $\alpha_2$ relaxations.}
\end{figure}

\textit{Mechanism for $\alpha_2$ process} - To gasp the microscopic origin of $\alpha_2$ process, we analysis the probabilities $p(u)$ of atomic displacements $u$ for Al and Sm atoms, respectively. As shown in Fig.~\ref{fig:fig4}(a), at a temperature $T = 500 K$ lower than the peak of $\alpha_2$ relaxations (about $560 K$), the peaks of $p(u)$ for Al and Sm separate with each other. While at the temperature higher than the $\alpha_2$ process, the $p(u)$ peaks of Al and Sm well overlaps with each other. This comparison implies that decoupling of the motions of Al and Sm atoms occurs when the temperature crosses the $\alpha_2$ peak.

To further quantify this behavior, Fig.~\ref{fig:fig4}(b) plots the most probable displacement, i.e. the peak position $u_p$ of $p(u)$ as a function of temperature. When the temperature increases from the lower regime, $u_p$ of Al and Sm atoms first increase separately until reaching a transition point where two curves merge to one. When comparing $u_p$ with loss moduli in Fig.~\ref{fig:fig4}(b), we find that the transition point coincides with the peak position of $\alpha_2$ relaxation over all the studied oscillation periods. Therefore, the $\alpha_2$ relaxation well correlates with the dynamical transition from coupling to decoupling motions of Al and Sm atoms. It indicates the asynchronous freezing of fast and slow motions could be the key factor leading to this process.

\begin{figure}
\includegraphics[width=0.42\textwidth]{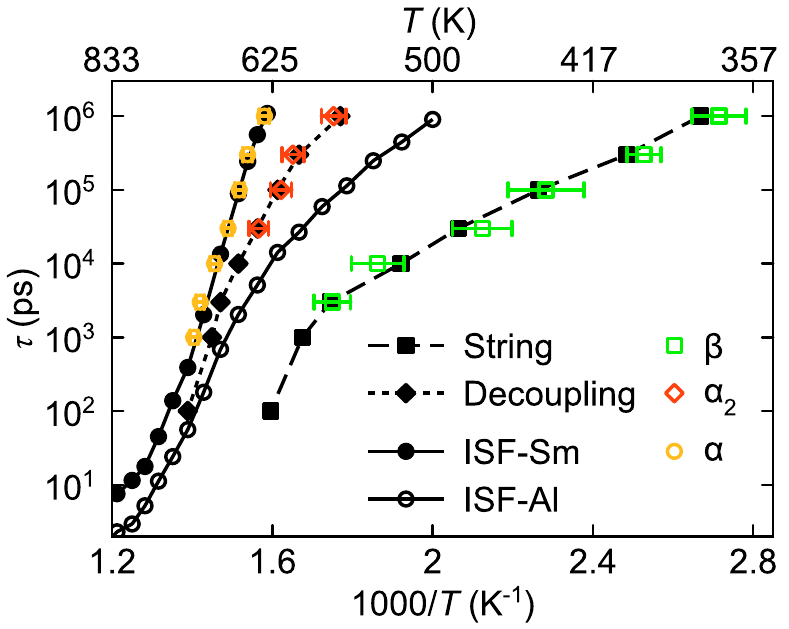}
\caption{\label{fig:fig5} Relaxation map of Al$_{\text{90}}$Sm$_{\text{10}}$ MG. ISF-Sm/Al: the $\alpha$-relaxation time of Sm/Al atoms based on the Sm/Al ISF spectra. Decoupling: the transition temperature from coupled to decoupled most probable motions $u_p (T)$ of Al and Sm atoms. String: the most probable temperature to form the long-string motions.}
\end{figure}

In Fig.~\ref{fig:fig5}, we summarized all the studied processes in a relaxation map over a wide range of temperature and timescales. One can see that the $\alpha_2$ relaxation and the decoupling between Sm and Al atoms follow the same temperature-time relation, suggesting an intrinsic correlation between them. Meanwhile, the $\beta$ relaxation and the long string-like motions ($\xi \geq 10$) agree with each other. Hence, the atomistic simulations not only predicted the complex relaxations in the MGs, but also elucidate the underlying mechanisms for them. 

Figure~\ref{fig:fig5} also reveals that the Al$_{\text{90}}$Sm$_{\text{10}}$ MG is a typical glass system in which the solute and solvent elements show dramatically different dynamical behaviors: the $\alpha$ relaxation time calculated based on the intermediate scattering function (ISF) of Sm atoms are orders of magnitudes longer than the Al atoms (see details in Supplemental Material \cite{Supp}). Moreover, the global $\alpha$ relaxation determined from MD-DMS correlates only with the $\alpha$ relaxation time from the ISF of Sm atoms, implying that it is controlled predominately by the slowest process.
Previous simulations \cite{Voigtmann2009,Cho2012, Bosse1995} suggested that the large atomic size ratio disparity might cause more than one glass transitions in model systems. In a recent theoretical work, Cui \textit{et al.} \cite{Cui2018} pointed out that the dynamical decoupling between constituents with wide mass disparity might lead to a separated relaxation process and suggested it to be a $\beta$ relaxation.
The identified $\alpha_2$ process here might be an experimental evidence for these scenarios in real glasses. Moreover, it indicates that the related process can be an additional primary process in stead of $\beta$ relaxation.
Finally, we note that the $\alpha_2$ process might not be unique to the Al$_{\text{90}}$Sm$_{\text{10}}$ MGs: in a recent work Xue \textit{et al.} \cite{Xue2015} reported the relaxation processes in a serials of LaGa-based MGs. Although not discussed explicitly, their data indeed exhibit a discernable $\alpha_2$-like process, which may also be related to the mobility decoupling between fast Ga and slow La atoms. 

We have shown that with atomic simulations, one could predict complex relaxation processes in MGs at the laboratory timescales and clarify their microscopic origins. A MG system with previously unidentified $\alpha_2$ relaxation process due to the mobility decoupling and strong $\beta$ relaxations caused by long-string motions has thus been predicted by simulations and verified by experiments. The combined experiments (validations) and simulations (predictions and clarification of mechanisms) represent a first glimpse of what may become a routine and integrated step in the study of glass relaxation. With above interpretations, one would expect an abundant $\alpha_2$ relaxations, or even more relaxation processes in  glass states. It then suggestions that efforts aimed at a quantitative theory to predict glass relaxation would be desirable. The results presented above thus open new challenges and opportunities for furthering our understanding of glass relaxations.

\begin{acknowledgments}
\textit{Acknowledgements} We thank Ms. Xiao-Hui Qin for experimental help. The work at HUST are supported by National Science Foundation of China (NSFC 51601064) and the Fundamental Research Funds for the Central Universities (2018KFYXKJC009). Work at Ames Laboratory was supported by the U.S. Department of Energy (DOE), Office of Science, Basic Energy Sciences, Materials Science and Engineering Division, under Contract No. DE-AC02-07CH11358, including a grant of computer time at the National Energy Research Supercomputing Center (NERSC) in Berkeley, CA.  
\end{acknowledgments}

\onecolumngrid
\bibliographystyle{apsrev4-1}

\begin{thebibliography}{50}
\bibitem{Berthier2011} L. Berthier and G. Biroli, Rev. Mod. Phys. \textbf{83}, 587 (2011).
\bibitem{Angell2000} C. A. Angell, K. L. Ngai, G. B. McKenna, P. F. McMillan, and S. W. Martin, J. Appl. Phys. \textbf{88}, 3113 (2000).
\bibitem{Micoulaut2016} M. Micoulaut, Rep. Prog. Phys. \textbf{79}, 066504 (2016).
\bibitem{Lunkenheimer2000} P. Lunkenheimer, U. Schneider, R. Brand, and A. Loid, Contemp. Phys. \textbf{41}, 15 (2000).
\bibitem{Sokolov2015} S. Khodadadi and A. P. Sokolov, Soft Matter \textbf{11}, 4984 (2015).
\bibitem{Pazmi2015} B. A. Pazmi{\~n}o Betancourt, P. Z. Hanakata, F. W. Starr, and J. F. Douglas, Proc. Natl. Acad. Sci. U.S.A. \textbf{112}, 2966 (2015).
\bibitem{Lahini2017} Y. Lahini, O. Gottesman, A. Amir, and S. M. Rubinstein, Phys. Rev. Lett. \textbf{118}, 085501 (2017).
\bibitem{Yu2014} H. B. Yu, W. H. Wang, H. Y. Bai, and K. Samwer, Natl. Sci. Rev. \textbf{1}, 429 (2014).
\bibitem{Yu2013} H.-B. Yu, W.-H. Wang, and K. Samwer, Mater. Today \textbf{16}, 183 (2013).
\bibitem{Ngai2004} K. L. Ngai and M. Paluch, J. Chem. Phys. \textbf{120}, 857 (2004).
\bibitem{Capaccioli2012} S. Capaccioli, M. Paluch, D. Prevosto, L.-M. Wang, and K. L. Ngai, J. Phys. Chem. Lett. \textbf{3}, 735 (2012).
\bibitem{Geirhos2018} K. Geirhos, P. Lunkenheimer, and A. Loidl, Phys. Rev. Lett. \textbf{120}, 085705 (2018).
\bibitem{Sondhaub2015} J. Sondhau{\ss}, M. Lantz, B. Gotsmann, and A. Schirmeisen, Langmuir \textbf{31}, 5398 (2015).
\bibitem{Welch2013} R. C. Welch, J. R. Smith, M. Potuzak, X. Guo, B. F. Bowden, T. J. Kiczenski, D. C. Allan, E. A. King, A. J. Ellison, and J. C. Mauro, Phys. Rev. Lett. \textbf{110}, 265901 (2013).
\bibitem{Kuchemann2017} S. Kuchemann and R. Maass, Scr. Mater. \textbf{137}, 5 (2017).
\bibitem{Luo2017} P. Luo, P. Wen, H. Y. Bai, B. Ruta, and W. H. Wang, Phys. Rev. Lett. \textbf{118}, 225901 (2017).
\bibitem{Wang2015} Q. Wang, S. T. Zhang, Y. Yang, Y. D. Dong, C. T. Liu, and J. Lu, Nat. Commun. \textbf{6}, 7876 (2015).
\bibitem{Yu2017} Y. Yu, M. Wang, M. M. Smedskjaer, J. C. Mauro, G. Sant, and M. Bauchy, Phys. Rev. Lett. \textbf{119}, 095501 (2017).
\bibitem{Cangialosi2013} D. Cangialosi, V. M. Boucher, A. Alegria, and J. Colmenero, Phys. Rev. Lett. \textbf{111}, 095701 (2013).
\bibitem{Ruta2012} B. Ruta, Y. Chushkin, G. Monaco, L. Cipelletti, E. Pineda, P. Bruna, V. M. Giordano, and M. Gonzalez-Silveira, Phys. Rev. Lett. \textbf{109}, 165701 (2012).
\bibitem{Wang2017} Q. Wang, J. J. Liu, Y. F. Ye, T. T. Liu, S. Wang, C. T. Liu, J. Lu, and Y. Yang, Mater. Today \textbf{20}, 293 (2017).
\bibitem{Ruta2017} B. Ruta, E. Pineda, and Z. Evenson, J. Phys. Condens. Matter \textbf{29}, 503002 (2017).
\bibitem{Bi2018} Q. L. Bi, Y. J. Lu, and W. H. Wang, Phys. Rev. Lett. \textbf{120}, 155501 (2018).
\bibitem{Yu2013a} H. B. Yu, K. Samwer, W. H. Wang, and H. Y. Bai, Nat. Commun. \textbf{4}, 2204 (2013).
\bibitem{Giordano2016} V. M. Giordano and B. Ruta, Nat. Commun. \textbf{7}, 10344 (2016).
\bibitem{Wang2016} B. Wang, B. S. Shang, X. Q. Gao, W. H. Wang, H. Y. Bai, M. X. Pan, and P. F. Guan, J. Phys. Chem. Lett. \textbf{7}, 4945 (2016).
\bibitem{Cheng2009} Y. Q. Cheng, E. Ma, and H. W. Sheng, Phys. Rev. Lett. \textbf{102}, 245501 (2009).
\bibitem{Cheng2011} Y. Q. Cheng and E. Ma, Prog. Mater. Sci. \textbf{56}, 379 (2011).
\bibitem{Falk2019}  Y. He, P. Yi, and M. L. Falk, Phys. Rev. Lett. \textbf{122}, 035501 (2019).
\bibitem{Mendelev2015} M. I. Mendelev, F. Zhang, Z. Ye, Y. Sun, M. C. Nguyen, S. R. Wilson, C. Z. Wang, and K. M. Ho, Model. Simul. Mater. Sci. Eng. \textbf{23}, 045013 (2015).
\bibitem{Sun2016} Y. Sun, Y. Zhang, F. Zhang, Z. Ye, Z. Ding, C.-Z. Wang, and K.-M. Ho, J. Appl. Phys. \textbf{120}, 015901 (2016).
\bibitem{Ye2017} Z. Ye, F. Zhang, Y. Sun, M. C. Nguyen, S. H. Zhou, L. Zhou, F. Meng, R. T. Ott, E. Park, M. F. Besser, M. J. Kramer, Z. J. Ding, M. I. Mendelev, C. Z. Wang, R. E. Napolitano, and K. M. Ho, Phys. Rev. Materials \textbf{1}, 055601 (2017).
\bibitem{Yang2018} L. Yang, F. Zhang, F.-Q. Meng, L. Zhou, Y. Sun, X. Zhao, Z. Ye, M. J. Kramer, C.-Z. Wang, and K.-M. Ho, Acta Mater. \textbf{156}, 97 (2018).
\bibitem{Wang2017a} L. Wang, J. Hoyt, N. Wang, and N. Provatas, arXiv \textbf{1810.05298} (2018).
\bibitem{Sun2019} Y. Sun, F. Zhang, L. Yang, H. Song, M. I. Mendelev, C.-Z. Wang, and K.-M. Ho, Phys. Rev. Materials \textbf{3}, 023404 (2019).
\bibitem{Zhao2018} L. Zhao, G. B. Bokas, J. H. Perepezko, and I. Szlufarska, Acta Mater. \textbf{142}, 1 (2018).
\bibitem{Yu2014a} H.-B. Yu and K. Samwer, Phys. Rev. B \textbf{90}, 144201 (2014).
\bibitem{Brown11} W. M. Brown, P. Wang, S. J. Plimpton, and A. N. Tharrington, Comp. Phys. Comm. \textbf{182}, 898 (2011).
\bibitem{Brown12} W. M. Brown, A. Kohlmeyer, S. J. Plimpton, and A. N. Tharrington, Comp. Phys. Comm. \textbf{183}, 449 (2012).
\bibitem{Brown13} W. M. Brown and Y. Masako, Comp. Phys. Comm. \textbf{184}, 2785 (2013).

\bibitem{Supp} See Supplemental Material [url] for technical details of simulations and experiments, the calculation details of string-like motions and the detailed intermediate scatter functions, which includes 
Refs. [42-44].

\bibitem{ye2017} Z. Ye, F. Zhang, Y. Sun, M. C. Nguyen, S. H. Zhou, L. Zhou, F. Meng, R. T. Ott, E. Park, M. F. Besser, M. J. Kramer, Z. J. Ding, M. I. Mendelev, C. Z. Wang, R. E. Napolitano, and K. M. Ho, Phys. Rev. Materials \textbf{1}, 055601 (2017).
\bibitem{kob2001} J. Horbach and W. Kob, Phys. Rev. E \textbf{64}, 041503 (2001).
\bibitem{simpleliquid} J.-P. Hansen and I. R. McDonald, Theory of Simple Liquids, 3rd ed. (Academic Press, London, 2006).


\bibitem{Yu2017a} H.-B. Yu, R. Richert, and K. Samwer, Sci. Adv. \textbf{3}, e1701577 (2017).
\bibitem{Yu2018} H.-B. Yu, M.-H. Yang, Y. Sun, F. Zhang, J.-B. Liu, C. Z. Wang, K. M. Ho, R. Richert, and K. Samwer, J. Phys. Chem. Lett. \textbf{9}, 5877 (2018).
\bibitem{Donati1998} C. Donati, J. F. Douglas, W. Kob, S. J. Plimpton, P. H. Poole, and S. C. Glotzer, Phys. Rev. Lett. \textbf{80}, 2338 (1998).
\bibitem{starr2013} F. W. Starr, J. F. Douglas, and S. Sastry, J. Chem. Phys. \textbf{138}, 12A541 (2013).
\bibitem{Zhang2015} H. Zhang, C. Zhong, J. F. Douglas, X. Wang, Q. Cao, D. Zhang, and J.-Z. Jiang, J. Chem. Phys. \textbf{142}, 164506 (2015).
\bibitem{Voigtmann2009} Th. Voigtmann and J. Horbach, Phys. Rev. Lett. \textbf{103}, 205901 (2009). 
\bibitem{Cho2012} H. W. Cho, G. Kwon, B.J. Sung, and A.Yethiraj, Phys. Rev. Lett. \textbf{109}, 155901 (2012).
\bibitem{Bosse1995} J. Bosse and Y. Kaneko, Phys. Rev. Lett. \textbf{74}, 4023 (1995).
\bibitem{Cui2018} B. Cui, Z. Evenson, B. Fan, M.-Z. Li, W.-H. Wang, and A. Zaccone, Phys. Rev. B \textbf{98}, 144201 (2018).
\bibitem{Xue2015} R. J. Xue, L. Z. Zhao, B. Zhang, H. Y. Bai, W. H. Wang, and M. X. Pan, Appl. Phys. Lett. \textbf{107}, 241902 (2015).
\end{thebibliography}

\clearpage


\begin{thebibliography}{11}
\bibitem{mys1} M. I. Mendelev, F. Zhang, Z. Ye, Y. Sun, M. C. Nguyen, S. R. Wilson, C. Z. Wang, and K. M. Ho, Model. Simul. Mater. Sci. Eng. \textbf{23}, 045013 (2015).
\bibitem{mys5} H.-B. Yu and K. Samwer, Phys. Rev. B \textbf{90}, 144201 (2014).
\bibitem{mys2} W. M. Brown, P. Wang, S. J. Plimpton, and A. N. Tharrington, Comput. Phys. Commun. \textbf{182}, 898 (2011).
\bibitem{mys3} W. M. Brown, A. Kohlmeyer, S. J. Plimpton, and A. N. Tharrington, Comput. Phys. Commun. \textbf{183}, 449 (2012).
\bibitem{mys4} W. M. Brown and M. Yamada, Comput. Phys. Commun. \textbf{184}, 2785 (2013).
\bibitem{ye2017} Z. Ye, F. Zhang, Y. Sun, M. C. Nguyen, S. H. Zhou, L. Zhou, F. Meng, R. T. Ott, E. Park, M. F. Besser, M. J. Kramer, Z. J. Ding, M. I. Mendelev, C. Z. Wang, R. E. Napolitano, and K. M. Ho, Phys. Rev. Materials \textbf{1}, 055601 (2017).
\bibitem{kob2001} J. Horbach and W. Kob, Phys. Rev. E \textbf{64}, 041503 (2001).
\bibitem{simpleliquid} J.-P. Hansen and I. R. McDonald, Theory of Simple Liquids, 3rd ed. (Academic Press, London, 2006).
\end{thebibliography}

\clearpage

\pagebreak
\widetext
\begin{center}
\textbf{\large Supplemental Material for ``Predicting Complex Relaxation Processes in Metallic Glass''} 
\end{center}

\twocolumngrid
\setcounter{equation}{0}
\setcounter{figure}{0}
\setcounter{table}{0}
\makeatletter
\renewcommand{\theequation}{S\arabic{equation}}
\renewcommand{\thefigure}{S\arabic{figure}}
\renewcommand{\bibnumfmt}[1]{[S#1]}
\renewcommand{\citenumfont}[1]{S#1}
\renewcommand{\thesection}{S\arabic{section}}

\setcounter{section}{0}
In this Supplemental Material we provide technical details of simulations and experiments in Sec. S1 and S2. In Sec. S3 we provide the calculation details of string-like motions and show the correlation between the peaks of $\beta$-relaxation and long string-like motions is independent to the threshold. In Sec. S4, we show the detailed intermediate scatter functions (ISF) for Al and Sm.

\section{Details of molecular dynamics simulation}
\begin{figure}
\includegraphics[width=0.45\textwidth]{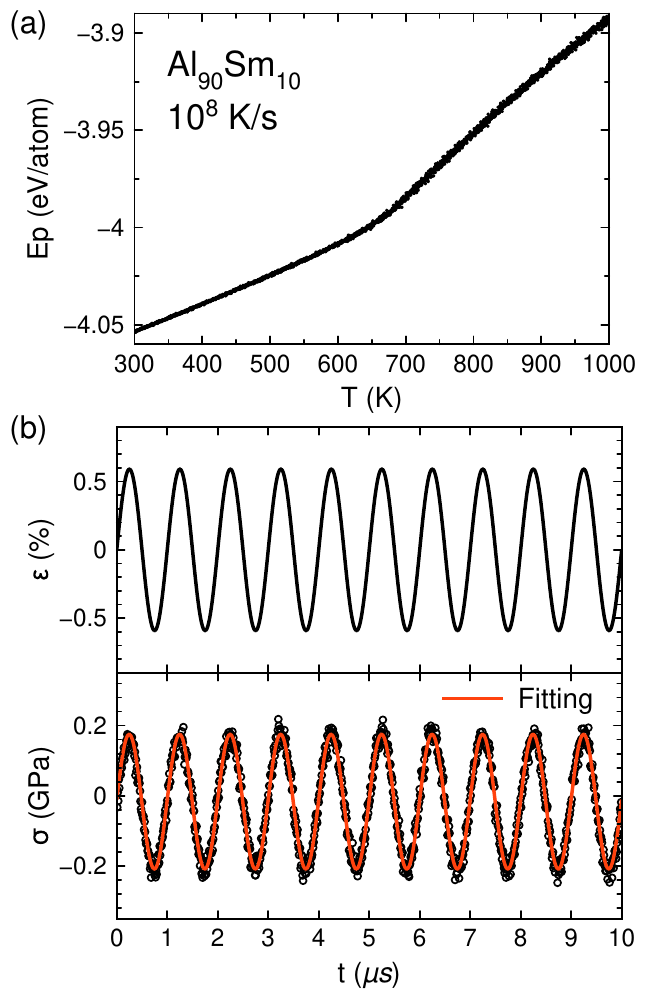}
\caption{\label{fig:figs1} (a) The potential energy as a function of temperature during the continuous cooling of the glass sample. (b) The strain $\varepsilon$ and measured stress $\sigma$ as a function of time for the glass model at $T=380K$. The time period $t_\omega=1\mu s$. The fitting function is $ \sigma (t) = \sigma_0 + \sigma_{A} \text{sin} ( 2 \pi t / t_{\omega} + \delta) $.  } \end{figure}

The molecular dynamics (MD) simulations were carried out using the Finnis-Sinclair potential \cite{mys1} with a 32,000-atom model of Al$_{\text{90}}$Sm$_{\text{10}}$. The Al$_{\text{90}}$Sm$_{\text{10}}$ melt were first well equilibrated at 1300K for 10 $ns$, followed by a continuous cooling to 200 K with a cooling rate $10^8$ K/s, using \textit{NPT} ensemble (that is, constant atom number, pressure, and temperature) and Nose-Hoover thermostat. The MD time step is 2 $fs$. The energy change during the cooling process is shown in Fig.~\ref{fig:figs1}(a), which indicates a clear glass transition.

The MD simulation of DMS (MD-DMS) \cite{mys5} was performed during the cooling process, covering a wide temperature range from the deeply undercooled liquid to low-temperature glass. Specifically, at a temperature $T$, we applied a sinusoidal strain $\varepsilon (t) = \varepsilon_A \sin (2 \pi t/t_\omega)$, with a oscillation period $t_\omega$ (related to frequency $f=1/t_\omega$) and a strain amplitude $\varepsilon_A$, along the $x$ direction of the metallic glass (MG) model. The resulting stress $\sigma (t)$ were measured and fitted to $\sigma (t) = \sigma_0 + \sigma_A \sin (2\pi t/t_\omega+\delta)$. $\delta$ is the phase difference between stress and strain. $\sigma_0$ is a linear term and usually small. One example is shown in Fig.~\ref{fig:figs1}(b). The strain was applied by a smooth change of the box length along the $x$ direction, whereas stresses were directly measured from each component of the pressure. From these values, storage ($E'$) and loss ($E''$) moduli are calculated as $E' =\sigma_A / \varepsilon_A \cos (\delta)$ and $E'' =\sigma_A / \varepsilon_A \sin (\delta)$, respectively. A strain amplitude $\varepsilon_A=0.6 \%$ was applied in all the MD-DMS, which ensured the deformations do not change the structure of the MG. The $NVT$ ensemble (constant number of atoms, volume, and temperature) were used during the MD-DMS simulations. 10 cycles were applied for each MD-DMS to measure the storage and loss moduli. All MD simulations were performed using the GPU-accelerated LAMMPS code \cite{mys2,mys3,mys4}.

\section{Details of Experiments}
\begin{figure}
\includegraphics[width=0.4\textwidth]{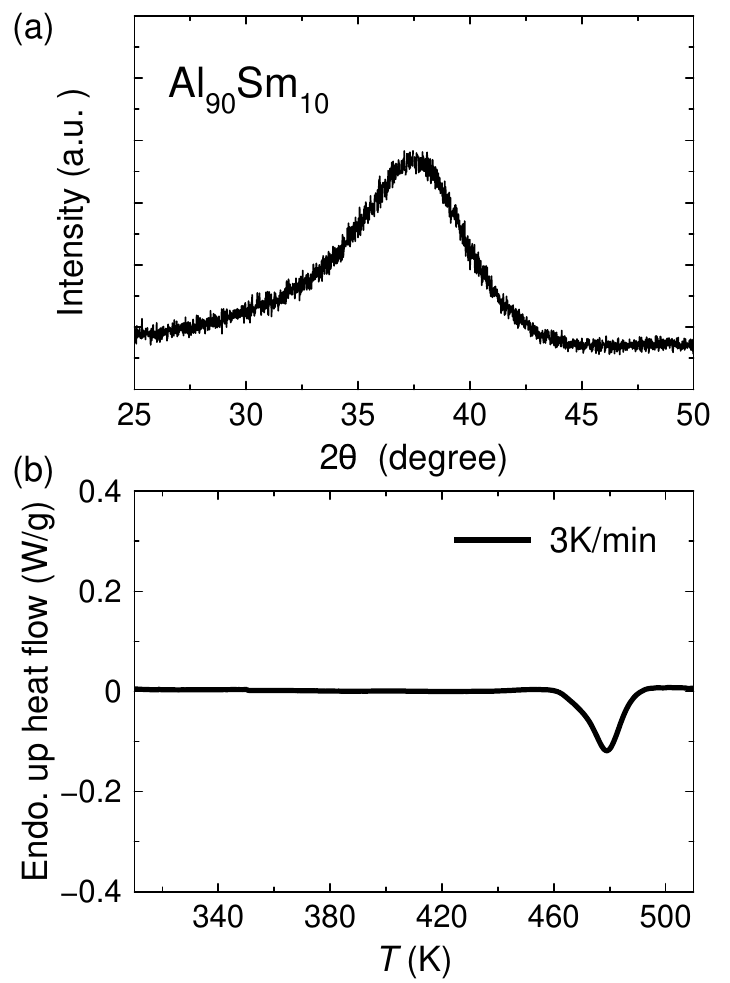}
\caption{\label{fig:figs2} (a) XRD patterns of Al$_{\text{90}}$Sm$_{\text{10}}$ MG ribbons. (b) DSC measurement with heating rate 3 K/min.} \end{figure}

\textit{Sample preparations} - The alloy ingot was prepared from a mixture of pure elements Al and Sm (purity $ \geq 99.99$ wt $\%$) by arc melting under the protection of high purity Argon atmosphere. In order to ensure the chemical homogeneity, the alloy was re-melted at least five times with magnetic stirring. The glassy ribbons were obtained via the melt-spinning. The amorphous nature of the ribbons was confirmed by X-ray diffraction (XRD, Bruke D2 phaser) which is shown in Fig. ~\ref{fig:figs2}(a).

\begin{figure}[b]
\includegraphics[width=0.85\textwidth]{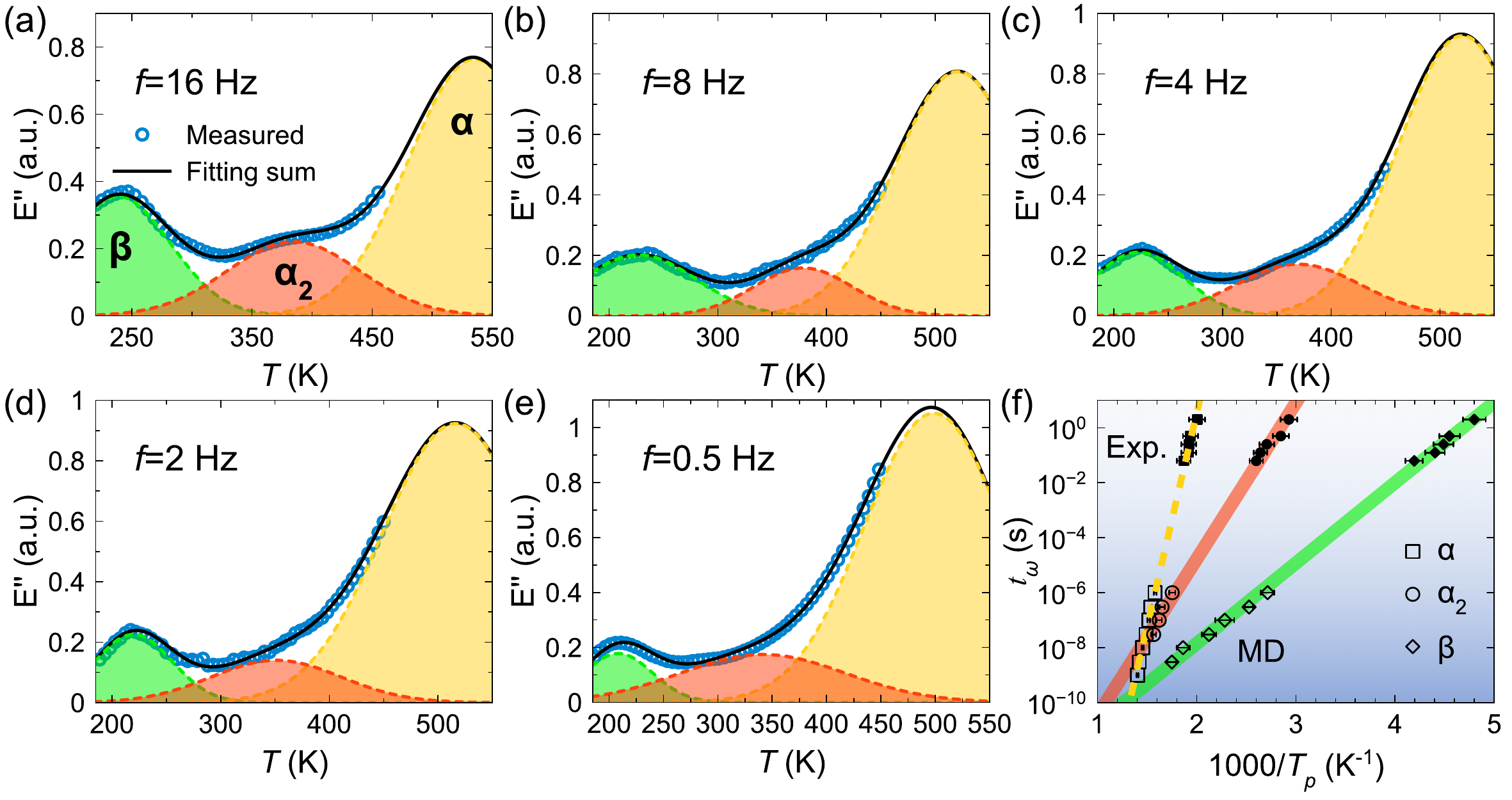}
\caption{\label{fig:figs3} (a)-(e) Experimental DMS fitted with three Gaussian peaks corresponding to the $\alpha$, $\alpha_2$ and $\beta$ relaxations, respectively. (f) The Arrhenius plot of fitted relaxation peak positions and oscillations periods for both experimental (solid symbols) and simulation (open symbols) data. The error bar is determined from the Gaussian fitting. The dashed line in (f) indicates the extrapolation from simulation condition to experimental condition for $\alpha$ peaks.}
\end{figure}

\textit{Dynamical mechanical analysis} - The relaxation dynamics of the MG was experimentally studied by a dynamical mechanical analyzer (DMA, TA Q800) using liquid Nitrogen for temperature control, which allows us to reach the cryogenic temperature (down to $150K$). The measurements were conducted during a temperature ramping of 3 K/min together with a film tension oscillation using the discrete testing frequencies of 0.5, 2, 4, 8 and 16 Hz. The measured loss moduli are fitted with three Gaussian functions as shown in Fig.~\ref{fig:figs3}. The peak temperatures of $\alpha_2$ and $\beta$ relaxation processes are compared with simulation results in Fig.~\ref{fig:figs3}(f), which shows a clear Arrehenius relation with the oscillation periods. Note the experimental $\alpha$ relaxation peaks cannot be directly fit to Gaussian functions because of the lack of data. To obtain a curve to guide eyes, we extrapolate simulation data to experimental frequencies to obtain initial guesses of peak positions. The peak shape is adjusted by fixing the fitting of $\alpha_2$ and $\beta$ peaks via the Levenberg-Marquardt nonlinear fitting algorithm implemented in the MagicPlot software (see \textit{Nonlinear Curve Fitting} in MagicPlot software manual, http://magicplot.com/wiki/fitting).

\textit{Calorimetry measurements} - To ensure the glassy state of samples during the DMA, we performed differential scanning calorimetry (DSC, Mettler -Toledo ADSC2) measurement with same heating rate in DMA. The measurements were performed under high purity nitrogen atmosphere at a flow rate of $50$ mL/min. To ensure the reliability of the data, the temperature and enthalpy scales were calibrated using the melting transitions of pure Indium and Zinc. In order to get a better accuracy, all the samples have a similar mass, about 12 mg. Each run was followed by a second run (on a fully crystallized sample) to obtain a baseline. As shown in Fig.~\ref{fig:figs2}(b), there is no heat flow in the temperature range of DMA experiments. However, the glass sample indeed devitrified at $ \sim 480 K $ as observed previously in Ref.\cite{ye2017}, which makes an upper limit temperature for DMA measurements.

\vfill

\onecolumngrid

\clearpage
\twocolumngrid

\section{String-like motions}
The string-like motions are recognized by analyzing the atomic rearrangement within the time interval of strain oscillation $t_\omega$. The displacement of atom $i$ is computed by $u_i=|{\boldsymbol{r}}_i(kt_\omega+t_\omega)-{\boldsymbol{r}}_i(kt_\omega)|$, where ${\boldsymbol{r}}_i(kt_\omega)$ is the atom position at the beginning of $k$-th deformation iteration and ${\boldsymbol{r}}_i(kt_\omega+t_\omega)$ is at the end of the deformation. This definition avoids any atomic displacement caused by the overall deformation of simulation box during MD-DMS. The atoms with $u>1.8 \text{\AA}$, i.e. the first minima of $p(u)$ as shown in Fig. 3(a) of the maintext, are defined as fast-moving atoms. If a fast-moving atom $i$ jumps to the initial position of another fast-moving atom $j$ after a deformation period, the atoms $i$ is defined as a string-like-moving atom. This can be express as $|{\boldsymbol{r}}_i(nt_\omega+t_\omega)-{\boldsymbol{r}}_j(nt_\omega)|<d_c$, where $d_c$ corresponds to the most probable displacement, i.e. the first peak position of $p(u)$. 

\begin{figure}
\includegraphics[width=0.35\textwidth]{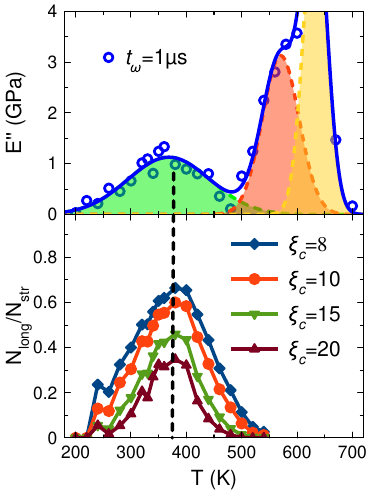}
\caption{\label{fig:figs4} Upper panel shows the peak of $\beta$ relaxation, lower panel shows the fraction of $\text{N}_{\text{long}} / \text{N}_{ \text{str} }$ with different $\xi_c$.  }
\end{figure}

In Fig.~\ref{fig:figs4}, we compute the ratio of atoms involved in the long-string motions $\text{N}_{\text{long}} / \text{N}_{ \text{str} }$ with different long-string threshold $\xi_c$. While the height of $\text{N}_{\text{long}} / \text{N}_{ \text{str} }(T)$ changes with the threshold, the peak position keeps constant. It clearlt indicates the correlation between $\beta$-relaxation and long string-like motions is independent to the threshold $\xi_c$.

\section{intermediate scattering function}
\begin{figure}[b]
\includegraphics[width=0.89\textwidth]{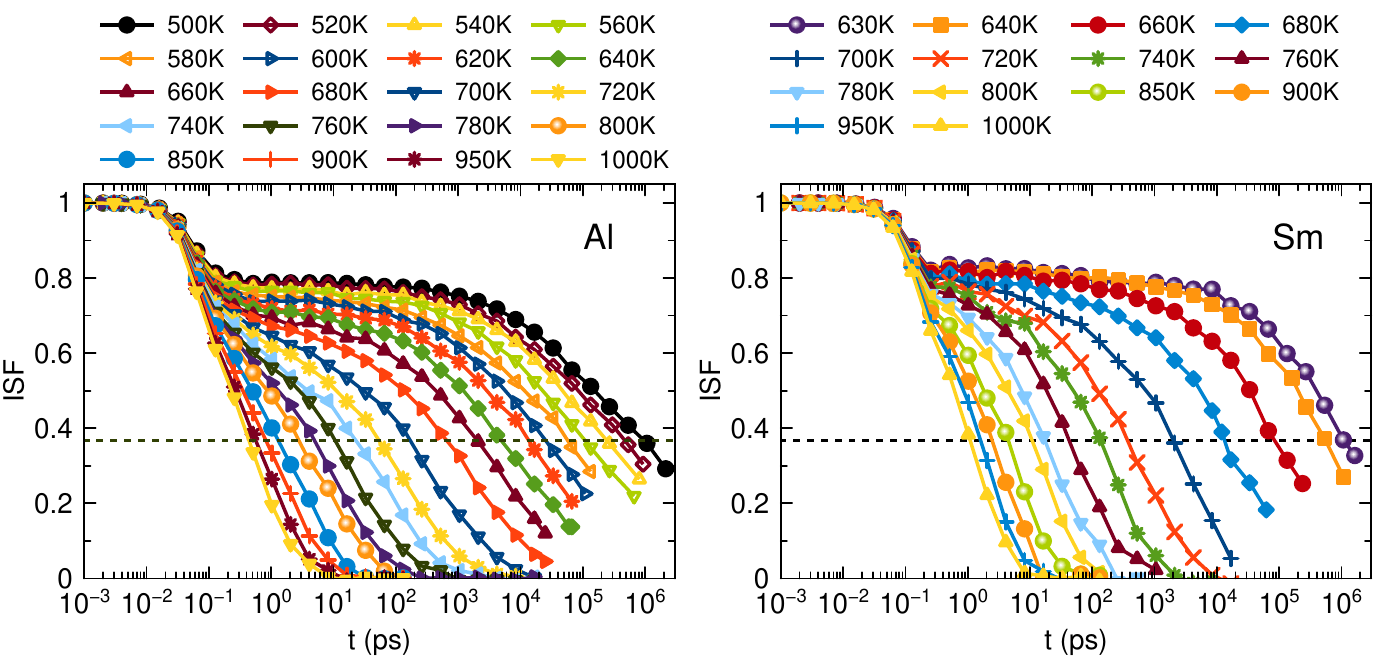}
\caption{\label{fig:figs5} The intermediate scatter functions (ISF) for Al and Sm in the Al$_{\text{90}}$Sm$_{\text{10}}$. }
\end{figure}

The self intermediate scattering functions (ISF) are computed for both Al and Sm atoms in the Al$_{\text{90}}$Sm$_{\text{10}}$ model by
\begin{equation}
F_s(\boldsymbol{Q},t)=\frac{1}{N} \sum_{j=1}^{N} \big< \text{exp}\{ i\boldsymbol{Q} \cdot [ \boldsymbol{r}_j(t) - \boldsymbol{r}_j(0)] \} \big>    
\label{eos}.
\end{equation}
To measure the $\alpha$-relaxation time,  the magnitude of wave vector $\boldsymbol{Q}$ is chosen as the first peak position of the structure factor $S(Q)$, which is $\sim 2.67 \text{\AA}^{-1}$ (with minor temperature dependence). The ISF results are shown in Fig.~\ref{fig:figs5}. The $\alpha$-relaxation time is obtained when the ISF decayed to $1/e$ \cite{simpleliquid, kob2001}.

\onecolumngrid

\end{document}